# Managing Blockchain Systems and Applications: A Process Model for Blockchain Configurations

Olga Labazova
University of Cologne
labazova@wiso.uni-koeln.de

Erol Kazan
IT University of Copenhagen
erka@itu.dk

Tobias Dehling
Karlsruhe Institute of Technology
dehling@kit.edu

Tuure Tuunanen
University of Jyväskylä
tuure.t.tuunanen@jyu.fi

Ali Sunyaev
Karlsruhe Institute of Technology
sunyaev@kit.edu

**Abstract.** *Blockchain is a radical innovation with a unique value proposition that shifts trust from institutions to algorithms. Still, the potential of blockchains remains elusive due to knowledge gaps between computer science research and socio-economic research. Building on information technology governance literature and the theory of coevolution, this study develops a process model for blockchain configurations that captures blockchain capability dimensions and application areas. We demonstrate the applicability of the proposed blockchain configuration process model on four blockchain projects. The proposed blockchain configuration process model assists with the selection and configuration of blockchain systems based on a set of known requirements for a blockchain project. Our findings contribute to research by bridging knowledge gaps between computer science and socio-economic research on blockchain. Specifically, we explore existing blockchain concepts and integrate them in a process model for blockchain configurations.*

## 1. Introduction

It will take years for blockchain systems to be fully adopted by businesses, but the journey has already begun (Iansiti and Lakhani, 2017). Blockchain is a radical innovation that has the potential to change the business logic for many industries. Blockchain's unique value proposition is the shift from institutional trust towards algorithmic consensus mechanisms (Beck and Müller-Bloch, 2017). Network nodes within blockchain systems process and record transactions within so-called blocks (Nakamoto, 2008). Blockchain systems have the advantage of removing single points of failure and improving data integrity and availability in contrast to centralized databases. To leverage the aforementioned features, organizations in different industries (e.g., finance, energy, healthcare) are considering to deploy blockchain systems to reduce intermediaries, decrease management costs, accelerate business processes, and tap into new revenue sources (Furlonger and Valdes, 2017).

However, current blockchain projects are more akin to trial-and-error approaches than purposeful information systems development due to a lack of best practices for blockchain development and unclear long-term business value (Furlonger and Valdes, 2017; Beck and Müller-Bloch, 2017). In other words, most initiated blockchain projects are prone for failure or inefficient resource allocations. For instance, ninety-two percent of 26,000 blockchain projects launched in 2016, with a total investment volume of over $1.5 billion, are defunct or indefinitely delayed (Trujillo, Fromhart and Srinivas, 2017).

The main reasons for failure are either flawed system designs or incompatible application areas (Risius and Spohrer, 2017). To illustrate, the Jasper project by the Bank of Canada revealed that a blockchain system for wholesale payments is not competitive compared to centralized systems with regards to its operating costs (Chapman et al., 2017). The example illustrates that narrow-scoped blockchain prototypes exhibit issues with regards to technical scalability, resource efficiency, user traceability, or lacking protection against fraud (Yli-Huumo et al., 2016; Xu et al., 2017).

The existing literature on blockchain focuses either on technical aspects or use cases (Lindman, Tuunainen and Rossi, 2017; Notheisen, Hawlitschek and Weinhardt, 2017). For instance, technical studies explore various consensus mechanisms and cryptographic protocols, predominately focusing on financial transactions as an application case (e.g., Bitcoin). On the other hand, research on blockchain use cases have their foci on business applications such as energy trading (Rutkin, 2016), healthcare (Azaria, Ekblaw, Vieira and Lippman, 2016), or supply chain management (Glaser, 2017; Mendling et al., 2017). However, the aforementioned studies are predominantly idea-driven and exhibit challenges due to a lack of feasible technical solutions (Avital et al., 2016; Lindman et al., 2017; Beck et al., 2017). To develop successful proof of concepts for blockchain application areas, research on blockchain technology (i.e., technical aspects) and blockchain applications (i.e., business use cases) should be considered conjointly.

We build on information technology (IT) governance literature and apply the theory of coevolution of technologies and application areas[1] (Grodal, Gotsopoulos and Suarez, 2015) as a lens to explore coevolving blockchain configurations and application areas. Bridging knowledge about the technology underlying blockchains and their application areas will create conditions for developing successful blockchain-based systems (Iansiti and Lakhani, 2017).

We explore the current body of blockchain research and present our model in the form of a process model for blockchain configurations that captures blockchain capabilities, that is, routinized, repeatable, and application-specific processes, enabling businesses to transform resources into business value (Ray, Muhanna and Barney, 2005; Tallon, 2007). We answer the following research question: *What application areas are advisable for blockchain systems and how can blockchain systems be purposefully configured across application cases?*

---
[1] By the term application areas we refer to the concept of categories as mentioned by Grodal et al. (2015).

To create the blockchain configuration process model, we systematically developed a taxonomy that groups blockchain application areas across mutually exclusive blockchain configurations (Nickerson, Varshney and Muntermann, 2013). The identified blockchain concepts and their relationships are consolidated in the blockchain configuration process model, which structures the blockchain concepts in four categories by semantic features and reciprocal relationships: (1) blockchain governance, (2) blockchain application areas, (3) blockchain properties, and (4) blockchain deployment. We illustrate the applicability of the proposed blockchain configuration process model on four selected blockchain projects.

With this research, we contribute to the extant research on blockchain by presenting a more granular and holistic view on identified blockchain concepts and their relationships. For practitioners, our proposed model offers guidance to managers to identify suitable blockchain systems and their corresponding application areas before development.

This paper proceeds as follows. In section 2, we discuss the theoretical aspects of IT governance and the theory of coevolution of technologies and applications. In section 3, we outline our three-step exploratory research approach for creating the blockchain configuration process model. In section 4, we present the blockchain configuration process model. In section 5, we illustrate the applicability of the blockchain configuration process model on four blockchain projects. In section 6, we discuss our findings, implications for theory and practice, and suggest avenues for future research.

## 2. Theoretical Background

### 2.1. IT Governance

IT governance can be defined as a collection of decision rights and accountabilities to encourage desirable behavior in the context of IT (Brown and Grant, 2005). Decision rights represent the governing control aspect over assets, whereas accountabilities capture the monitoring of decision-making processes. Incentives play a vital part in IT governance, because they motivate and guide agents to act favorably for the purposes of specific systems. Overall, the literature on IT governance discusses three basic governance approaches. First, centralized governance includes executive committees for decision-making and is characterized by having centralized business processes, providing control over architectures, and possessing formal assessments and monitoring decisions. Second, a decentralized approach to IT governance requires no or few governance mechanisms for decision-making and insists on local accountabilities (Brown and Magill, 1994; Schwarz and Hirschheim, 2003; Brown and Grant, 2005). Lastly, companies that aim to balance the benefits of the centralized and decentralized models follow a hybrid governance approach. These companies establish a centralized group to provide core services while allowing business units to control a portion of the overall functions (Boynton and Zmud, 1987; Rockart, 1988).

### 2.1.1. Blockchain Governance

The most successful blockchain systems will be those that adapt their governance to the organizational environments for business value creation (Kharitonov, 2017; Beck et al., 2018). Being introduced as a more or less decentralized data management solution, blockchain systems evolve continuously and are aligned with different IT governance approaches. Beck et al. (2018) specify decision rights as a dimension of blockchain centralization. Decision-making power can either be concentrated in few governing nodes or distributed equally among all nodes in the blockchain network. With regard to accountabilities, they differ in their rights to monitor decisions on blockchain systems, having the ability to adjust actions based on consequences incurred (Beck et al., 2018). In the same vein, different incentive schemes motivate agents to act within blockchain systems for monetary or non-monetary rewards.

## 2.2. Theory of Co-evolution of Technologies and Applications

The theory of coevolution of technologies and application areas during industry emergence focuses on mechanisms of their continuous coevolution, which starts with a period of divergence and continues with a period of convergence (Grodal et al., 2015). The period of divergence is characterized by high diversity in technology to address emerging application requirements. Technologies evolve and fulfill more application requirements through continuous design recombination. Application areas are influenced by a pool of ready-made technological designs, which in turn satisfy groups of application requirements. The following period of convergence results in consensus among producers with respect to efficient technological designs for mature application areas.

The blockchain domain is currently at an early stage of industry emergence and is characterized by a high diversity of technological designs and potential application areas (Lindman et al., 2017; Miscione, Ziolkowski, Zavolokina and Schwabe, 2018; Schlegel, Zavolokina and Schwabe, 2018). A variety of consensus mechanisms (Karame, Androulaki and Capkun, 2012) and anonymity schemes (Reid and Harrigan, 2013) produce various experimental solutions that are largely unrelated or opaque to emerging blockchain application cases (Yli-Huumo et al., 2016). Nevertheless, the number of blockchain application experiments are growing, leading to different blockchain-based services, such as supply chain management (Glaser, 2017; Mendling et al., 2017), energy trading (Rutkin, 2016), or authentication services (Miscione et al., 2018).

In turn, blockchain application cases are not fully supported by ready-made technological solutions (Risius and Spohrer, 2017). So far, extant research on blockchain systems yields isolated and unstructured concepts and offers only limited support for configuring blockchain systems for application areas.

## 3. Research Approach

Our research approach for developing the blockchain configuration process model comprises three consecutive steps (Figure 1). First, we explore blockchain concepts and their relationships through taxonomy development based on literature, business reports, and instantiated decentralized applications (Nickerson et al., 2013). Second, we structure the findings in the form of the blockchain configuration process model. Third, we illustrate the applicability of the blockchain configuration process model on four blockchain projects.

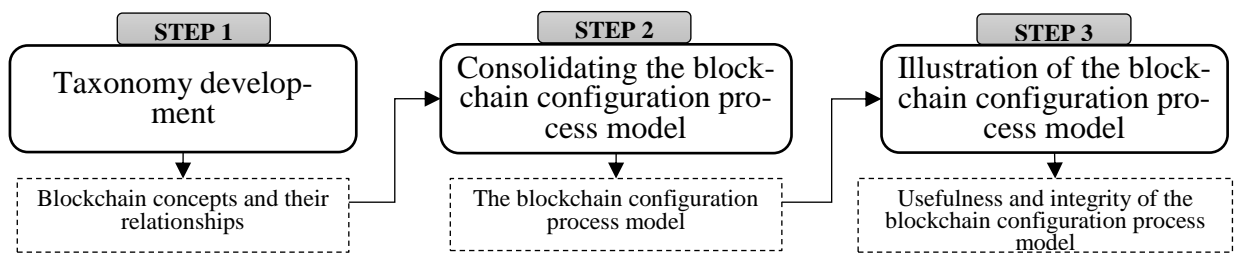

**Figure 1.** Research Approach. The Blockchain Configuration Process Model

### 3.1. Taxonomy Development

To organize extant knowledge on blockchain, we employed the taxonomy development method proposed by Nickerson et al. (2013), who define a taxonomy as a set of dimensions. Each dimension consists of "mutually exclusive and collectively exhaustive characteristics in a way that each object under consideration has one and only one characteristic in every dimension" (Nickerson et al., 2013, p.5). The taxonomy development method proceeds in three stages. In the initial stage, a metacharacteristic and ending conditions are defined according to the purposes of the taxonomy to be developed. In the main stage, the taxonomy is developed. Objects to be classified with the taxonomy (in this study, application cases, dimensions, and characteristics) are identified during inductive or deductive iterations. During inductive iterations, empirical cases are analyzed to determine dimensions and characteristics for the taxonomy. During deductive iterations, dimensions and characteristics are derived from the existing scientific knowledge base. In the final stage, the taxonomy is evaluated against ending conditions.

The aim of the taxonomy is to derive and classify blockchain application areas and dimensions driven by blockchain characteristics. Therefore, we selected blockchain characteristics (e.g., consensus mechanism, anonymity level) as a metacharacteristic. The metacharacteristic serves as basis for identification of further dimensions and characteristics.

We developed the taxonomy in four iterations. The first three iterations were inductive iterations, where we identified application cases to derive dimensions and characteristics. For each inductive iteration, we used different types of sources: scientific literature, business reviews, and white papers of blockchain applications, respectively. The fourth iteration was a deductive iteration where we revised the taxonomy based on previous classifications of blockchain systems. *In the first iteration*, we searched papers and articles in the web of science core collection[2] with the search string "blockchain OR distributed ledger" on October 17, 2016, in title, abstract, and keywords, covering the whole period of publications (Webster and Watson, 2002; Vom Brocke et al., 2009). The search returned fifty-one papers and articles. After screening the titles and abstracts, we discarded ten papers as non-blockchain research and coded the forty-one remaining relevant papers and articles. In the first iteration, we identified six dimensions with fourteen characteristics and six application areas with ten application cases. The analysis of the existing scientific literature revealed detailed information on separate blockchain characteristics (e.g., consensus mechanisms) or specific blockchain application examples (e.g., energy markets, prediction platforms) but lacked comprehensiveness.

*In the second iteration*, we analyzed business reports, which provide less precise, but more comprehensive information. We investigated twenty business reports published by national agencies, consulting companies, and international institutions. We revised the taxonomy and added two dimensions, six characteristics, and one application case.

To fill the remaining gaps in the taxonomy, we reviewed eighty-six blockchain systems and applications (e.g., Bitcoin, Ethereum, Hyperledger) *in the third iteration.* If possible, we used the applications; otherwise, we read available documentation and white papers. During the third iteration, we added four new application cases.

*The fourth iteration* was deductive, where we derived characteristics, dimensions, and application cases from fifteen previous classifications. We used all previous classifications that we could identify in extant literature until October 2018. This analysis showed that our taxonomy is consistent with extant blockchain classifications.

All ending conditions proposed by Nickerson at al. (2013) were fulfilled after the fourth iteration. First, all found blockchain application cases described in the existing scientific literature or business reports can be classified with the taxonomy. Second, each dimension is unique

---

[2] Used indices: "Science Citation Index Expanded (1900-present), Social Sciences Citation Index (1900-present), Arts & Humanities Citation Index (1975-present), Conference Proceedings Citation Index- Science (1990-present), Conference Proceedings Citation Index- Social Science & Humanities (1990-present), Book Citation Index– Science (2005-present), Book Citation Index– Social Sciences & Humanities (2005-present), and Emerging Sources Citation Index (2015-present)"

and mutually exclusive and each characteristic is unique within its own dimension. Third, all application cases were classified with a single characteristic for each dimension. Fourth, the taxonomy is concise—consists only of dimensions that classify application cases. Fifth, the taxonomy is robust—differentiates each application case from all others. Sixth, the taxonomy is explanatory, comprehensive, and extensible—highlights the main features of each application case and can be extended when new application cases arise.

### 3.2. Consolidation of the Findings

Based on the taxonomy development, we synthesized the findings into a process model for blockchain configurations. The model captures characteristics and application areas that are pertinent to blockchain systems. Specifically, the model is structured by four dimensions, which are distinct by their semantic features and reciprocal relationships: (1) blockchain governance, (2) blockchain application area, (3) blockchain properties, and (4) blockchain deployment. To synthesize blockchain concepts and investigate their relationships, we coded the data using three types of coding schemes—open coding, axial coding, and selective coding[3] (Strauss and Corbin, 1990). We applied open coding for initial categorization of blockchain concepts; axial coding for removal of overlapping concepts while iteratively testing the blockchain concepts against the data, and selective coding to identify the relationships between concepts. One researcher coded the sources three times (November 2016, April 2017, November 2017) and another researcher validated the results after each iteration (Strauss, 1987). Disputes were resolved in group discussions.

## 4. The Blockchain Configuration Process Model

Based on a set of known requirements of a blockchain project (i.e., blockchain governance, blockchain application area), the blockchain configuration process model (Figure 2) supports configuration of blockchain properties and selection of blockchain deployment attributes (i.e., processing and settlement of transactions).

The blockchain configuration process model proceeds in three steps. First, one chooses the suitable governance approach (decentralized, hybrid, or centralized) and the application area (i.e., financial transactions, enforcements, asset management, storage, communication, or ranking) that reflects the requirements of the blockchain project. Examples of blockchain applications are located at the intersection of blockchain governance and application area. Second, the proposed model identifies appropriate blockchain properties according to the selected application area (e.g., financial transactions). The blockchain properties are token (equity, utility), customizability (no, fixed, custom), data type (logs, assets), and history retention (whole, updates). Third, the blockchain configuration process model supports the blockchain deployment (i.e., processing and settlement) according to the selected blockchain governance approach (e.g., decentralized). The blockchain deployment attributes comprise access (i.e., private, public), validation (i.e., permissioned, unpermissioned), consensus mechanism (i.e., proof-of-work, proof-of-stake, Practical Byzantine Fault Tolerance, self-developed consensus mechanism), and the anonymity level (i.e., anonymous, pseudonymous, identifiable).

---

[3] Open coding is a process for grouping categories and subcategories (Strauss & Corbin, 1990, p.12). Axial coding is a process for testing "that categories are related to their subcategories, and the relationships against data" (Strauss & Corbin, 1990, p.13). Selective coding is a process "by which all categories are unified around a 'core' category, and categories that need further explication are filled-in with descriptive details" (Strauss & Corbin, 1990, p.14).

After finishing the three steps, the blockchain configuration process model terminates. For complex blockchain projects that include different blockchain capabilities, the process can be reiterated.

### 4.1. Blockchain Governance

The blockchain configuration process model accounts for different approaches of IT governance—decentralized, hybrid, and centralized (Brown and Grant, 2005). *A decentralized approach to blockchain governance* implies that all nodes in the network have decision rights and accountability rights. Bitcoin is an example of blockchains with decentralized governance. In the Bitcoin network, all participants hold rights to decide on the correct functioning of the system, whereas transparency of the data on blockchains allows all actors to monitor decisions (Nakamoto, 2008). Collectively governed companies or startups often require decentralized blockchain governance to spread decision rights and accountabilities among all actors in the network to reduce the network overload.

**I. CHOOSE THE BLOCKCHAIN GOVERNANCE. CHOOSE THE APPLICATION AREA.**

| Application Area \ Blockchain Governance | Decentralized | Hybrid | Centralized |
|---|---|---|---|
| **Financial Transactions** | Cryptocurrencies; Wealth Storage; Micro-Payments | Cross-Border and Micro-Interorganizational Payments | Central-Issued Financial Instruments |
| **Enforcements** | Enforcements between Individuals | Interorganizational Enforcements | Central-Issued Enforcements |
| **Asset Management** | Authentication; Ownership; Audit Trails; Access Management | Interorganizational Asset Management | Enterprise Asset Management |
| **Storage** | Decentralized Storage | Not Explored for Blockchain Systems | |
| **Communication** | Messengers; IoT Communication | | |
| **Ranking** | Reputation; Rating | | |

**II. CHOOSE THE BLOCKCHAIN PROPERTIES FOR THE APPLICATION AREA**

| Application Area \ Blockchain Properties | Token | Customizability | Data Type | History Retention |
|---|---|---|---|---|
| **Financial Transaction** | Equity | No | Logs | Whole |
| **Enforcement** | Utility | Custom | Logs | Whole |
| **Asset Management** | Utility | Fixed | Logs | Whole |
| **Storage** | Utility | Fixed | Assets | Updates |
| **Communication** | Utility | Fixed | Assets | Whole |
| **Ranking** | Utility | Fixed | Logs | Updates |

**III. CHOOSE THE BLOCKCHAIN DEPLOYMENT ATTRIBUTES FOR THE BLOCKCHAIN GOVERNANCE**

| Blockchain Governance \ Blockchain Deployment | Access | Validation | Consensus Mechanism | Anonymity Level |
|---|---|---|---|---|
| **Decentralized** | Public | Unpermissioned | Proof-of-Work/ Proof-of-Stake | Anonymous/ Pseudonymous |
| **Hybrid** | Private | Unpermissioned | Practical Byzantine Fault Tolerance | Identifiable |
| **Centralized** | Private | Permissioned | Self-Developed Consensus | Identifiable |

**Figure 2.** The Blockchain Configuration Process Model

Blockchains that are governed by a *hybrid governance approach* allow only authenticated and predefined users to monitor decisions. However, ones a node is a part of the network, participation in decision-making requires no additional permissions. Ripple is an example of a blockchain with hybrid governance. In the Ripple network, predefined nodes are trusted organizations that deal directly with each other to support a peer-to-peer financial settlement system (Walsh et al., 2016). A hybrid approach to blockchain governance is useful for inter-organizational collaboration, where blockchains keep the network closed to ensure the confidentiality of the information, whereas the decision rights are distributed among all nodes in the network.

*A centralized approach to blockchain governance* supports blockchains where nodes (usually only a small number of nodes) that have been authorized to validate transactions require additional authorization to have decision rights. An example for systems with centralized blockchain governance are IBM (Hyperledger) blockchains, which support regulatory and supervisory nodes to monitor the system (Hyperledger Architecture Working Group, 2017). Centralized blockchain governance is useful to support enterprise business projects, where a predefined number of users in the network, usually semi-trusted organizations or individuals, can monitor decisions while only few nodes have rights to validate transactions.

### 4.2. Blockchain Application Areas

The taxonomy yields six blockchain application areas, which comprise a total of fourteen application cases. Application areas group application cases with similar semantic features, for instance, usage scenarios, and with similar combinations of blockchain configurations. The first application area is *financial transactions*, which captures application cases concerned with money transfer and exchange. Anonymous and conventional cryptocurrencies, wealth storage, and micro-payments utilize blockchains with decentralized governance (e.g., Bitcoin, Primecoin, Namecoin, Zcash, and Darkcoin). Interorganizational cross-border and micro-financial transactions employ a hybrid approach to blockchain governance (e.g., Ripple, Stellar). Central-issued financial instruments are deployed on blockchains with centralized governance. For instance, RSCoin and Fedcoin projects (Koning, 2016) may allow federal states to independently launch coins.

The second application area of blockchains is *enforcement*. Enforcements ensure compliance with laws, regulations, rules, standards, or social norms through application logic (Beck et al., 2018). Blockchain-based enforcements between individuals can, for instance, be developed on the Ethereum platform, which supports a decentralized approach to blockchain governance. Interorganizational enforcements usually employ blockchains with hybrid governance, such as Ripple Codius, which allows executing enforcements between predefined organizations. Blockchains with centralized governance can be useful for deployment of centrally issued enforcements (e.g., R3 Corda). For example, UK Barclays Bank built a prototype on the R3 Corda platform that translates legal contracts into smart contracts, where all involved parties can monitor (but not decide) on amendments to the original smart contracts (Walsh et al., 2016).

The third application area is *asset management* and concerned with management tasks such as authentication, know your customer services, luxury goods provenance, and control of business assets. The management of off-chain registered assets usually requires decentralized governance of blockchains. For example, a user can prove the ownership or verify the origin of an asset by keeping their labels on the Bitcoin blockchain (e.g., Colored Coins). To keep information confidential, interorganizational asset management (e.g., Everledger) applies hybrid blockchain governance. For enterprises, blockchains with centralized governance may be suitable for managing interorganizational assets. For example, Maersk and IBM introduce TradeLens, a platform for real-time access to shipping data and shipping documents that utilizes a Hyperledger-based blockchain.

The fourth application area is *storage* and is concerned with keeping digital assets, such as certificates or music and video files, on blockchains (Kishigami et al., 2015). Blockchain-based decentralized storage is implemented on blockchains with decentralized governance, because it requires a high number of nodes to distribute the transaction load of the network. For instance, the Storj project leverages sharding to split encrypted data (Wilkinson et al., 2014). Blockchains with hybrid and centralized governance seem inappropriate for blockchain-based storage, because of extensive resource requirements for blockchain deployment and the availability of effective alternative solutions, such as decentralized storage services (Salviotti, de Rossi and Abbatemarco, 2018).

The fifth application area is *communication*. Messaging and IoT communication can be realized on blockchains with decentralized governance, because the content is intended for mass communication (e.g., Whisper; Unger et al., 2015). Communication systems based on blockchains with hybrid and centralized governance do not create additional value compared to peer-to-peer messengers such as Telehash, which are used by many decentralized services (e.g., IBM Adept).

The sixth application area is *ranking* with a single application case. Global reputation & rating (e.g., Dennis and Owenson, 2016) is supported by blockchains with decentralized governance and allows a number of untrusted participants to create blockchain-based reputations. Blockchains with hybrid and centralized governance seem inappropriate for ranking, because of the availability of alternative solutions, for example, ranking based on peer-to-peer systems such as Gnutella (Salviotti et al., 2018).

### 4.3. Blockchain Properties

Blockchain properties allow for configuration of blockchains according to application areas. We identify four important blockchain properties. *Token* specify how transactions processed by a blockchain are represented. Equity tokens capture transfer of value between parties (e.g., Alice transfers 1 Bitcoin to Bob). Utility tokens are more elaborate and contain more extensive data and application logic. *Customizability* captures a blockchain's ability to process application logic. No customizability indicates that the blockchain cannot handle application logic. Fixed customizability supports built-in configurations. If customizability is custom, blockchains support processing of application logic provided by users. *Data type* focuses on the type of data shared between blockchain users. Logs imply an exchange of logs of executed transactions. Digital assets mean that the whole digital assets such as documents, messages, and video or music files, are exchanged. *History retention* ascertains whether the whole blockchain, starting with a genesis block, or only its recent updates are kept and distributed between nodes.

The choice of blockchain properties depends on blockchain application areas. Everything that is primarily used for *financial transactions* is based on equity tokens. Financial transactions require no customizability. For example, the Bitcoin scripting language is purposefully not Turing-complete (Walsh et al., 2016). Blockchains for financial transactions exchange logs of executed financial transactions and keep the whole transaction history (Nakamoto, 2008).

*Enforcements* are based on utility tokens (e.g., company stock ownership) (Beck et al., 2018). Enforcements are customized by smart contracts, which are executed across participants in the blockchain network (Peters, Panayi and Chapelle, 2015). Enforcements exchange logs of smart contracts and retrieve the whole history of logs for security reasons.

*Asset management* is based on utility tokens (e.g., data access). Fixed customizability allows users to use built-in configurations. Executed actions under the assets are represented by logs (e.g., access to assets, asset changes), which are continuously kept on the blockchain.

Blockchain-based *storage* is supported by utility tokens and provides for fixed customizability of blockchains. Storage blockchains keep digital assets. To improve scalability of blockchains, only recent updates of assets are stored. Users are more interested in the current state of the assets and not in their changes over time.

*Communication* employs utility tokens, fixed customizability of blockchains, and exchange of digital assets in the form of text messages. Blockchains keep the whole communication history (Pureswaran, Panikkar, Nair, & Brody, 2015; IBM Adept). However, application cases are far from the production stage.

*Blockchain Application in Information Systems Research**Ranking* uses utility tokens and allows for fixed customizability. Blockchains exchange logs of actions, usually reputation or rating scores, and keeps only recent updates of the transaction history, because outdated votes are not necessary for calculating reputation or rating and can be safely removed from blockchains (Dennis and Owenson, 2016).

### 4.4. Blockchain Deployment

Blockchain deployment attributes depend on blockchain governance. We identified four blockchain deployment attributes. *Access* represents the ability to read and submit data on a blockchain (Beck et al., 2018). Private access makes a blockchain available for reading and submitting data only to authorized users. Public access allows everyone to read data from and submit data to a blockchain. *Validation* indicates different mechanisms for validating transactions on a blockchain. Permissioned validation means that only authorized users validate transactions and participate in consensus finding. If validation is unpermissioned, all users in the network validate transactions. *Consensus mechanism* is concerned with mechanisms for reaching consensus on blockchain updates. Proof-of-work requires validating notes to spend resources (or work), usually processor time or storage space. Proof-of-stake requires users to proof the ownership of tokens to establish their stake in the blockchain. Practical Byzantine Fault Tolerance requires agreement by the majority of validating nodes (2/3 of validating nodes) for transaction validation. Self-developed consensus mechanisms usually include several highly trusted nodes for achieving system-level agreements. *Anonymity level* assesses with what accuracy users can be matched to particular identities. If the characteristic is anonymous, users do not have to provide any identifying information to work with a blockchain. If the characteristic is pseudonymous, users have to work under a pseudonym. Blockchains with the characteristic identifiable ask for or automatically collect personally identifiable information such as email addresses or IP addresses.

A *decentralized approach to blockchain governance* implies public access to blockchains, which allows all participants in the network to monitor transactions. Unpermissioned validation invites all participants to participate in consensus finding. Proof-of-work or proof-of-stake consensus mechanisms ensure the correct functioning of the blockchain system in a network with a large number of untrusted nodes. Blockchains with decentralized governance support anonymity and pseudonymity of users.

A *hybrid approach to blockchain governance* requires private access to blockchains that makes a blockchain only available to authorized users. However, unpermissioned validation requires all users in the blockchain network to participate in consensus finding. The blockchain network consists of a small number of trusted nodes that makes it possible to use energy-efficient but communication-heavy Practical Byzantine Fault Tolerance as consensus mechanism. Blockchains with hybrid governance frequently ask for name, surname, and email address (e.g., Hyperledger, Ripple) to make users identifiable.

In blockchains with *centralized governance*, private access allows only authorized users to monitor transactions. Permissioned validation allows only authorized users to validate transactions and participate in consensus finding. Often validating nodes find consensus based on resource-saving, self-developed mechanisms. Nodes in private-permissioned blockchains must be identifiable and trusted.

## 5. Four Blockchain Projects

We illustrate the applicability and usefulness of the blockchain configuration process model on four blockchain projects: (1) DB Systel & IBM (public mobility), (2) LitSonar (academic literature tool), (3) dSCM Tool (data sharing between factories), and (4) Blockchain4openscience.org (portal for researchers).

We selected different types of blockchain projects that exhibit three different blockchain governance approaches (i.e., centralized, hybrid, and decentralized) and different application areas to demonstrate its analytical capabilities for identifying common and dissimilar configurations. We conducted four open-ended, semi-structured interviews with leading researchers, solution architects, or leading developers in September and October 2018. Interviews lasted between 46 and 85 minutes with an average duration of 58 minutes. The interviews were transcribed and coded using NVivo software. During the interviews, interviewees applied the blockchain configuration process model to their projects and discussed the usefulness of the blockchain configuration process model, blockchain concepts, and their relationships according to their blockchain project. In addition, we used secondary data sources to triangulate data and understand the relationship between blockchain concepts and actual use. We gathered 229 pages of interview transcriptions and secondary data (Table 1).

| Table 1. Project-by-Project Evaluation. Data Collection | | | | | |
|---|---|---|---|---|---|
| | | **Interviews** | | | **Secondary Data** | |
| № | Project | Interviewee | Time (min.) | Transcription (pages) | Sources | Number of pages |
| 1 | DB Systel & IBM: Public Mobility | IBM Solution Architect | 46 | 9 | ibm.com, hyperledger.com, Hyperledger white papers | 120 |
| 3 | dSCM Tool | Leading Developer | 51 | 11 | ethereum.org, scientific sources | 53 |

| | | | | | |
|---|---|---|---|---|---|
| 4 | LitSonar | Leading Researcher | 85 | 15 | litsonar.com, scientific sources | 33 |
| 5 | Blockchain4open science.org | Leading Researcher | 48 | 9 | blockchain4openscience.com, scientific sources | 23 |
| | Overall | | 230 | 58 | Overall | 229 |

We demonstrate how the choice of blockchain governance, blockchain application area, blockchain properties, and blockchain deployment are consistent and interrelated (Table 2).

| **Table 2.** Overview of Blockchain Projects | | | | |
|---|---|---|---|---|
| № | **Project** | **Blockchain Governance** | **Application Area** | **Blockchain Properties** | **Blockchain Deployment** |
| 1 | DB Systel & IBM: Public Mobility | Centralized | Financial Transactions | Equity token, fixed customizability, logs, whole history retention | Private access, permissioned validation, Practical Byzantine Fault Tolerance, identifiable nodes |
| 2 | dSCM Tool | Hybrid | Communication | Utility & equity tokens, custom customizability, logs, whole history retention | Public & private access, unpermissioned & permissioned validation, proof-of-work, pseudonymous nodes |
| 3 | LitSonar | Decentralized | Storage | Utility tokens, fixed customizability, digital assets, recent updates | Public access, unpermissioned validation, proof-of-existence, pseudonymous nodes |
| 4 | Blockchain4openscience | Decentralized | Ranking | Utility tokens, fixed customizability, logs exchange, whole history retention | Public access, validation is not defined, consequence-based consensus, identifiable nodes |

### 5.1. DB Systel & IBM: Public Mobility

DB Systel GmbH (Frankfurt, Germany) is a digital research and innovation business unit within the Deutsche Bahn Group, one of the largest global mobility and logistics companies in Europe (IBM, 2018). As a strategic partnership between IBM and Deutsche Bahn, DB Systel GmbH helps to develop a public mobility solution using blockchain.

**Blockchain Governance.** The blockchain is *centrally governed* by IBM. The interviewed IBM solution architect stated: *"IBM does not have its own blockchain implementation. We have built-in IBM add-ons for governance of the blockchain network on top of Hyperledger open source licenses."*

**Blockchain application area.** The primary focus of the application are financial transactions. DB Systel GmbH wants to provide a single ticket for journeys with different mobility providers. The application enables mobility providers to check the validity of tickets and reimburses them for services rendered in a transparent and immutable way.

**Blockchain properties.** Tokens in the application represent tickets. Accordingly, they can be considered equity tokens since their main purpose is to represent the monetary value of the ticket. Tokens offering additional features were not considered by the project. The financial transactions *are not customized*. However, additional business logic is enforced by smart contracts integrated into the architecture and triggered by ticketing transactions (IBM, 2018). Hence, the application offers fixed customizability. The blockchain *logs* all transactions performed with the application. The interviewed IBM solution architect confirmed: "*We build solutions of having transactions where the data is small and does not contain, for example, the process-sensitive data.*" The application stores *the whole transaction history* to keep a record of all transactions performed on the system. The interviewed IBM solution architect specified: "*If you want to have a blockchain for years and if your asset is critical, then you need the history.*"

**Blockchain deployment.** The approach to blockchain governance (i.e., centralized governance) reveals the following blockchain deployment attributes. Access to blockchain is *private* and available only for authorized users. Mobility providers have an additional (*permissioned*) authentication for being validating members in the blockchain network and get the split of the revenue. The interviewed IBM solution architect explained: "*Users register to an application. Parties, who provide services for the tickets have to be participants in the blockchain, or to trust one of the participants*". A variant of Practical Byzantine Fault Tolerance is employed as consensus mechanism. The IBM solution architect confirmed: "*As you are working with private-permissioned blockchain, you can do other consensus protocols, which allow you to be simple and more efficient in validation.*" Users are identifiable within the application. The interviewed IBM solution architect specified: "*IBM provides blockchain for business, for business means that we are looking for a way to have blockchains with identities.*"

### 5.2. dSCM Tool

The dSCM tool is a part of a larger project that aims to provide data sharing in federation clouds between different factories in supply chain management networks, which record and store data from sensors. Each factory and supplier have access to the data in different stages of aggregation.

**Blockchain governance.** The system *employs a hybrid blockchain governance* approach. The system is built on the public Ethereum blockchain so that transactions are visible to the general public. The leading developer confirmed: "*We rather focus on the level of openness than the level of trust. If we want to have a higher level of trust, we just would set up the network in a different way, with some auditing or monitoring mechanisms*". However, the system also integrates the HAWK protocol to delegate sensitive smart contracts to managers. Managers are third-parties that all participants in the respective smart contract trust. Hence, the governance approach constitutes a hybrid form where the Ethereum's decentralized governance approach is extended with an additional layer to maintain confidentiality of sensitive information.

**Blockchain application area.** The blockchain application area is communication. The focus of the system is to enable *communication* and transparent data exchange between the factories in the supply chain network.

**Blockchain properties.** *Utility tokens* are employed, because the system has to track interactions beyond simple value transactions. However, equity tokens are used as well to pay for smart contract execution. The leading developer explained: *"We use both tokens, equity tokens because of this gas, and utility tokens, because of additional features that are offered, like storing text or smart contracts inside."* The system provides users with an integrated development environment for arbitrary smart contracts to realize additional *built-in business logic*. Hence, customizability is custom. The system keeps *logs of executed actions* and holds *the whole history* of the actions for security reasons.

**Blockchain deployment.** The tool employs the Ethereum platform, because it allows for fast deployment of a proof of concept. Ethereum supports *public access* to the data on blockchain and *unpermissioned validation*. The consensus mechanism is the most challenging part to select while deploying blockchains. The leading developer stated: *"I have no idea about the consensus mechanism that would be the best. I just stayed with proof-of-work."* The nodes in the system are organizations that want to protect sensitive information from competitors in the network. Currently, the developers are exploring how to keep the parties involved in interactions and the content of transactions confidential from third-parties. To this end, security technologies like mixers or the HAWK protocol are employed.

### 5.3. LitSonar

The LitSonar project aims to establish a repository for open science/open access publications. Aims of the project are to make open access publications and data sets available to the general public and to ensure their authenticity.

.........

**Blockchain governance.** The project is going to use a public-unpermissioned blockchain to assure high availability and integrity by allowing for as many validating nodes as possible. Hence, the governance approach is *decentralized.*

**Blockchain application area.** As infrastructure for open *storage*[4] of the scientific data, the project considers using blockchain. The leading researcher confirmed: *"Storage is the biggest issue for the project. We believe that blockchain might be the best option to implement an infrastructure that allows everyone to access the knowledge."* The blockchain will most likely serve as trust anchor and facilitator for access management. The Interplanetary File System is currently considered as a decentralized storage medium.

**Blockchain properties.** The application employs utility tokens, because representing scientific literature requires more information than simple transactions transferring monetary values. Customizability is fixed since users are allowed to execute *fixed features* (e.g., data selection, commenting, and uploading). The blockchain exchanges *digital assets*. The system will capture only *recent updates* of the assets to avoid effort for management of outdated information.

**Blockchain deployment.** The project requires *public access* to the blockchain data. The leading researcher explained: *"It (the application) will be one big database where everyone can publish everything."* The system employs *unpermissioned validation* to enable any interested party to contribute to the system. However, scientific libraries are the most likely candidates for operating validating nodes. The considered consensus mechanism is *proof-of-existence*, a proof-of-work algorithm that builds on provision of storage capacities instead of computational resources. The leading researcher stated: *"You need a working blockchain that allows you to make proof-of-existence, giving timestamps for documents to search these verified indexes, which contain verified documents."* Users are pseudonymous. The leading researcher explained: *"Anonymity cannot establish links between papers and users, while identifiability creates privacy concerns."*

### 5.4. Blockchain4openscience.org

The open-source project Blockchain4openscience.org aims to represent extended information about research and researchers—their scholarly works, research interests, and affiliations.

---

[4] The project is at an early stage and struggles with the technical issues (e.g., scalability) to store more than a pointer to the data but the actual data.

**Blockchain governance.** The project uses a public-unpermissioned blockchain, because everyone should have rights to create and monitor ledgers of the scientific reputation. Accordingly, the governance is *decentralized*.

**Blockchain application area.** Blockchain4openscience.org implements blockchains for open science in collaboration with openvivo.org, an existing social platform for managing scientific reputation. The project uses various reward mechanisms for the scientific works to generate the scientific *reputation* (Garcia, Lopez and Conlon, 2018). The leading researcher explained: *"Once you have links where data is stored, you can facilitate the mechanism to reach the data and establish relationships across data. By doing that, you generate tokens that are stored in a wallet of the scientific reputation"*.

**Blockchain properties.** The project selected *a utility token* to avoid generating an economy around data. The leading researcher explained: *"The reputation-based blockchain; the incentive is not money."* The blockchain should support *fixed events* such as token generation and rely on *the whole history* of the transactions[5]. The exchanged data is *logs*, while the real data is stored off-chain. The leading researcher confirmed: *"Yes, right now we are experimenting with IPFS, where we have the actual things for the blockchain that are represented with a hash."*

**Blockchain deployment.** The access to blockchain should be *public*. The leading researcher explained: *"You have all the assets and basically what you want is to make the data available and public."* The project team had not decided on an approach for transactions validation. The system uses *consequence-based consensus*, which is based on certificated quoting. The users in the system are fully *identifiable,* because anonymity does not support managing reputation. Besides, the project uses Open Researcher and Contributor ID (ORCID).

## 6. Discussion

### 6.1. Principal Findings

Based on a method for taxonomy development for deriving blockchain system characteristics and dimensions (Nickerson et al., 2013), the blockchain configuration process model represents an analytical tool to assists with the selection and configuration of blockchain systems based on known requirements (i.e., blockchain governance, blockchain application area). The proposed model is particularly useful for decision makers to derive insights before initiating the development of a blockchain project, and allows the classification of alternative blockchain configurations to identify performance differences.

---

[5] The concept of rolling blockchain was introduced. To keep recent updates, the miners should add a checkpoint to a block, in which all blocks older than some point in time can be safely removed (Dennis and Owenson, 2016).

In this study, we derived four groups of relationships between blockchain concepts, which assist with the development of blockchain-based systems and consolidates extant knowledge on the blockchain domain (Figure 3).

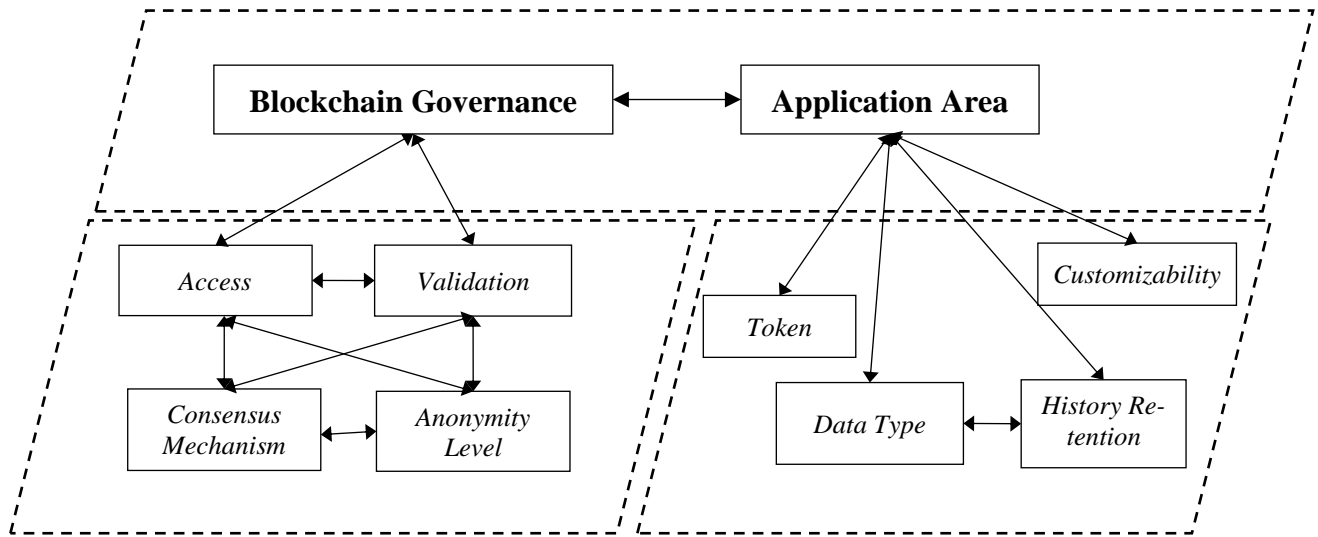

**Figure 3.** Overview of Blockchain Concepts and their Relationships

6.1.1. Relationships between Blockchain Governance and Application Areas

Our exploratory study revealed interdependencies between blockchain governance and blockchain application areas. First, not all blockchain governance approaches are well suited for all application areas. For instance, the application areas financial transactions, enforcements, and asset management can benefit from all three blockchain governance approaches. However, application areas such as communication system, decentralized storage, and ranking should be performed on blockchains with a decentralized governance approach, because operating such systems is resource intensive and more effective solutions all already available if lesser degrees of centralization are satisfactory (e.g., ranking portals operated by a third party).

6.1.2. Relationships between Application Areas and Blockchain Properties

The findings suggest that blockchain properties (i.e., token, customizability, data type, and history retention) have reciprocal relationships with application areas. To illustrate, equity *tokens* support financial transactions, while other application areas require tokens with more functionality; thus, they employ utility tokens. *Customizability* is not required for financial transactions since all required operations can be clearly specified upfront. For enforcements are customizability is custom by design, because smart contracts must be specified. Other application areas employ customizability and provide users with built-in configuration options (e.g., a set of different types of transactions). *Data type* has the property logs for the application areas financial transactions, enforcements, data management, and ranking since only logs of executed actions are exchanged and stored. The application areas storage and communication keep digital assets directly in the blockchain instead of logging state transitions. *History retention* is influenced by application areas and data type depending on sensitivity or volume stored in a blockchain. If larger amounts of data are to be stored in a blockchain, only recent updates can be retained to prevent the blockchain from requiring too much storage space. Limited retention of transaction history is also an option to alleviate confidentiality challenges.

### 6.1.3. Relationships between Blockchain Governance and Blockchain Deployment

There are relationships between *blockchain governance* and the attributes regarding *access* and *validation*. A decentralized blockchain governance approach requires public access and unpermissioned validation, because the network can be monitored and set up by all participating nodes. A hybrid blockchain governance approach supports private access and unpermissioned validation, because authorization is required for accessing sensitive data (see dSCM Tool). Lastly, centralized blockchain governance requires private access to blockchain systems, because the nodes need to be authorized to read data and perform permissioned validation (Hyperledger Architecture Working Group, 2017).

### 6.1.4. Relationships between Blockchain Deployment Attributes

To support efficiency, trade-offs between blockchain deployment attributes of *access* and *validation* (public vs. private; permissioned vs. unpermissioned) determine the level of blockchain decentralization (Walsh et al., 2016; Beck et al., 2018). Further, combinations in the attributes of *access* and *validation* determine the *consensus mechanism* and the degree of anonymity. In blockchains with public access and unpermissioned validation (e.g., Ethereum), the large number of untrusted nodes require proof-of-work or proof-of-stake consensus mechanisms. Blockchains with private access and unpermissioned validation have a smaller number of trusted nodes; hence, it is feasible to employ the energy-saving but communication-heavy Practical Byzantine Fault Tolerance algorithm. In blockchain systems with private access and permissioned validation, highly trusted nodes participate in consensus finding based on self-developed rules. The degree of *anonymity* is influenced by access and validation type due to different requirements for authentication and authorization schemes; for instance, in a permissioned blockchain users cannot be anonymous by design. Only blockchain systems with public access and unpermissioned validation (e.g., Ethereum) support anonymity or pseudonymity, because no authentication is required.

Moreover, *consensus mechanisms* and the degree of *anonymity* influence each other. Proof-of-work and proof-of-stake support anonymous and pseudonymous nodes. On the other hand, Practical Byzantine Fault Tolerance and self-developed consensus mechanisms/rules require the identification of users to ensure control and trust in private blockchain systems.

### 6.2. Theoretical Contributions

This study contributes to research in four ways. First, our findings contribute to the theory of IT governance in the context of blockchain systems based on three IT governance approaches. A decentralized blockchain governance is suitable for collectively governed companies or startups to spread decision rights and accountabilities among all actors (i.e., nodes) in the network. A hybrid blockchain governance is useful for interorganizational collaboration to ensure confidentiality while sharing decision rights among all nodes in the network. Lastly, a centralized blockchain governance approach is useful to support enterprise business projects where a predefined number of users can monitor decisions while only validating nodes have decision rights.

Second, previous research on blockchain proposes concepts of interest to computer science (Yli-Huumo et al., 2016) and business applications related concepts (Salviotti et al., 2018), but falls short in exploring their relationships. Our study bridges prior research by offering clearer conceptualizations for the identified concepts and their relationships. As such, the identified relationships (i.e., betwee governance and application areas) bridge knowledge gaps between the computer science and the socio-economic literature on blockchain.

Third, the identified blockchain properties (i.e., token, customizability, data type, and history retention) are important attributes for configuring blockchain systems, which, however, have received only little attention in the literature.

Last, we establish an overview of blockchain application cases and more abstract blockchain application areas. Application areas group application cases with similar characteristics (e.g., usage scenarios) and similar combinations of blockchain characteristics.

### 6.3. Practical Contributions

Our research contributes to practice in three ways. First, our proposed blockchain configuration process model guides design of blockchain-based information systems prior to development. The model establishes an overview of best practices for blockchain applications and organizes them according to their configurations. This is particularly useful for practitioners to identify promising blockchain projects and assess risks before their implementations.

Second, we highlight the hybrid governance approach, besides the widely-known decentralized or centralized ones. For many application cases, organizations may consider blockchain implementations with hybrid governance that store information in a more confidential but still more decentralized fashion (e.g., Ethereum with the Hawk framework).

Third, we provide an overview of application cases and areas beyond the financial sector, which is still the focus of the majority of current blockchain projects. For example, the entertainment industry may use blockchain-based data management to monitor the use of media content for billing purposes and to prevent copyright infringements.

### 6.4. Limitations

This study is not without limitations. First, due to our focus on blockchain governance (Beck et al., 2018) and coevolution of technology and application areas (Grodal et al., 2015), we do not consider other socio-economic concepts; therefore, blockchain concepts and their relationships can be enriched with different socio-economic concepts.

Second, the identified application areas do not directly capture more complex services such as prediction markets or crowdsourcing platforms; instead, we decided to break complex application cases down into the basic functionalities that can be performed by blockchains.

Third, we identified several application areas proposed in scientific and business sources, which would be invalid within the scope of the blockchain configuration process model (e.g., private blockchains that use proof-of-work consensus mechanism and, thus, unnecessarily waste resources). Such examples were purposefully disregarded for design of the blockchain configuration process model, because they seem unreasonable for commercial use.

### 6.5. Future research

There are three promising areas for future research. First, exploratory research could replicate our three-step research approach with different scientific and business sources to falsify or corroborate our findings. In addition, the identification of further blockchain configurations and application areas would broaden the applicability of the blockchain configuration process model. For example, the addition of a dimension data structure with the two characteristics blocks and graph would make the blockchain configuration process model also applicable to ledgers based on directed acyclic graphs (e.g., Otte, de Vos and Pouwelse, 2020) that may replace certain blockchain architectures in the future for some application cases, for example, in the internet-of-things with its scarce resources.

Second, research focused on other socio-economic concepts, for example, market regulations in different countries or implementation and management strategies of blockchain-based information systems will be useful to contextualize the blockchain configuration process model for different industries and domains. For example, interoperability of blockchain-based systems and other information systems is constrained by industry-specific or country-specific data protection regulations such as HIPAA in healthcare (Mercuri, 2004; Sunyaev, Leimeister, Schweiger and Krcmar, 2008) or the GDPR in the European Union (The European Parliament and The European Council, 2016). Third, studies in different industry contexts would allow to development measurements and performance indicators that are pertinent for blockchain systems. This, in turn, would reduce the existing uncertainty about the real business value of blockchain systems (Notheisen et al., 2017).

## 7. Conclusion

Blockchain is an emerging technology with a largely untapped potential for enhancement of many aspects in the information systems domain. Currently, research streams on blockchain remain largely disconnected, which prevents further development of blockchain industries and hinders integration of blockchain-based information systems into the business landscape. Based on theory of IT governance and the theory of coevolution of technologies and application areas during industry emergence (Grodal et al., 2015), our work consolidates knowledge on blockchain governance, application areas, blockchain properties, and blockchain deployment attributes in the form of the blockchain configuration process model. This research contributes to literature by clarifying the concept of blockchain governance, summary of generic blockchain application areas, and highlighting new concepts that complement existing blockchain literature. Overall, the blockchain configuration process model captures blockchain capabilities based on the current state of knowledge on blockchain. Simultaneously, it serves as a foundation for future research of blockchain integrations into the business landscape.

## 8. Acknowledgements

We would like to thank Elke Kunde (IBM), Mike Conlon (University of Florida), and Alexander García (Graz University of Technology) for the agreement to participate in research.